\documentclass[12pt]{article}
\usepackage{epsfig}
\usepackage{psfrag}
\usepackage{amssymb, amsmath}
\usepackage{latexsym}
\def\ps {SU(4)_C \times SU(2)_L \times SU(2)_R}

\def\be{\begin{equation}}
\def\ee{\end{equation}}
\def\bc{\begin{center}}
\def\ec{\end{center}}
\def\bea{\begin{eqnarray}}
\def\eea{\end{eqnarray}}

\def\nn{\nonumber}

\catcode`@=11
\def\marginnote#1{}
\newcount\hour
\newcount\minute
\newtoks\amorpm
\hour=\time\divide\hour by60
\minute=\time{\multiply\hour by60 \global\advance\minute by-\hour}
\edef\standardtime{{\ifnum\hour<12 \global\amorpm={am}%
        \else\global\amorpm={pm}\advance\hour by-12 \fi
        \ifnum\hour=0 \hour=12 \fi
        \number\hour:\ifnum\minute<10 0\fi\number\minute\the\amorpm}}
\edef\militarytime{\number\hour:\ifnum\minute<10 0\fi\number\minute}
\def\draftlabel#1{{\@bsphack\if@filesw {\let\thepage\relax
   \xdef\@gtempa{\write\@auxout{\string
      \newlabel{#1}{{\@currentlabel}{\thepage}}}}}\@gtempa
   \if@nobreak \ifvmode\nobreak\fi\fi\fi\@esphack}
        \gdef\@eqnlabel{#1}}
\def\@eqnlabel{}
\def\@vacuum{}
\def\draftmarginnote#1{\marginpar{\raggedright\scriptsize\tt#1}}
\def\draft{\oddsidemargin 0.0truein
        \def\@oddfoot{\sl preliminary draft \hfil
        \rm\thepage\hfil\sl\today\quad\militarytime}
        \let\@evenfoot\@oddfoot \overfullrule 3pt
        \let\label=\draftlabel
        \let\marginnote=\draftmarginnote
   \def\@eqnnum{(\theequation)\rlap{\kern\marginparsep\tt\@eqnlabel}%
\global\let\@eqnlabel\@vacuum}  }
\catcode`@=12
\begin{document}
\begin{titlepage}
\vspace*{-1cm}

\vskip 0.5cm
\begin{center}
{\Large\bf 
Fermion masses and proton decay in a minimal five-dimensional SO(10) model
}
\end{center}
\vskip 0.5  cm
\begin{center}
{\large Maria Laura Alciati}~\footnote{e-mail address: maria.laura.alciati@pd.infn.it}
~~~{\large Ferruccio Feruglio}~\footnote{e-mail address: feruglio@pd.infn.it}
~~~{\large Yin Lin}~\footnote{e-mail address: yinlin@pd.infn.it}
\\
\vskip .1cm
Dipartimento di Fisica `G.~Galilei', Universit\`a di Padova
\\
INFN, Sezione di Padova, Via Marzolo~8, I-35131 Padua, Italy
\\
\vskip .2cm
{\large Alvise Varagnolo}~\footnote{e-mail address: a.varagnolo@sns.it}
\\
\vskip .1cm
Scuola Normale Superiore, Pisa
\\
INFN, Sezione di Pisa, I-56126 Pisa, Italy
\\
\end{center}
\vskip 0.7cm
\begin{abstract}
\noindent 
We propose a minimal SO(10) model in 5 space-time dimensions. 
The single extra spatial dimension is
compactified on the orbifold $S^1/(Z_2\times Z_2')$ reducing the
gauge group to that of Pati-Salam $\ps$.
The breaking down to the standard model group is obtained
through an ordinary Higgs mechanism taking place
at the Pati-Salam brane, giving rise to a proper
gauge coupling unification.
We achieve a correct description of fermion masses and mixing angles
by 
describing first and second generations as bulk fields,
and by embedding the third generation into
four multiplets located at the Pati-Salam brane.
The Yukawa sector is simple and compact and
predicts a neutrino spectrum of normal hierarchy type.
Concerning proton decay, dimension five operators are absent
and the essentially unique localization of matter multiplets
implies that the minimal couplings between the super-heavy gauge
bosons and matter fields are vanishing. Non-minimal interactions 
are allowed but the resulting dimension six operators describing
proton decay are too suppressed to produce observable effects, 
even in future, super-massive detectors.
\end{abstract}

\end{titlepage}
\setcounter{footnote}{0}
\vskip2truecm

%
%
\section{Introduction}
Despite the absence of any direct experimental check, the idea of grand 
unification is so deeply influencing our present view of particle physics
that it has become a standard ingredient of most of the constructions
extending the Standard Model (SM). 
It is however a matter of fact that all the advantages of grand unification,
such as gauge coupling unification, particle classification,
charge quantization,
are quite difficult to incorporate into a complete and consistent
picture. All simple realizations based on the standard tools of 
four-dimensional (4D) quantum field theory (QFT) face severe problems like the 
doublet-triplet (DT) splitting problem, a too fast proton decay, 
wrong fermion mass relations and unsatisfactory gauge 
coupling unification beyond the leading order.
Non-minimal 4D versions of GUTs exist, which offer solutions to some or all the
above mentioned problems, but very often these constructions
make use of elaborate epicycles that spoil the beauty of the
original GUT ideas, in order to be viable \cite{afm}. Quite often the
necessity of these complicated constructions arises from the 
highly non trivial sector
needed to successfully break the GUT symmetry, to naturally produce
the DT splitting and to correctly break the flavour symmetries
of the theory. 

These difficulties have eventually led to the idea that in 
nature, perhaps, grand unification is not realized 
as a conventional 4D QFT. Examples 
of non conventional realizations of grand unification can
be found in the context of string theory where, in some
circumstances, the GUT symmetry
becomes manifest only in the presence of the complete
string spectrum. Working at the level of QFTs, successful
versions of GUTs have been formulated in a higher dimensional
space-time \cite{kawa,af1,hebecker1,hebecker2,hn1,hn2}. Indeed, going to higher
dimensions, several GUT problems find simple and elegant solutions \cite{kawa}.
For instance in SU(5), by allowing a single compact extra dimension, 
the GUT symmetry can be efficiently
broken down to the SM gauge group by the compactification mechanism,
without including any Higgs multiplet. Moreover this reduction of the
gauge symmetry automatically entails a DT splitting, as soon as the 
5 and $\bar{5}$ Higgs representations are introduced as 5D fields.
Also gauge coupling unification, analyzed at the next-to-leading order
can benefit \cite{hn1,hn2} from this framework 
\footnote{The accompanying theoretical 
uncertainty is however similar to the one of conventional 4D constructions.}. The size of the required extra dimension(s) is tiny
since its inverse sets the grand unified scale. Thus the extra space
does not have a direct impact on the gauge hierarchy problem. However it 
can highly affect our description of flavour physics because
Yukawa couplings are strictly related to the localization
properties of matter in the extra compact space \cite{flavour}.

Higher-dimensional GUTs make specific predictions about proton decay,
which represents the ultimate experimental test of grand unification.
For instance, in 5D supersymmetric (SUSY) SU(5), 
dimension 5 operators arising through
coloured higgsino exchange are forbidden by an abelian continuous 
symmetry,
thus avoiding the dominant and problematic source of proton decay 
in 4D SUSY models. Proton decay is dominated by dimension six
operators due to the exchange of the heavy gauge boson X, whose mass
corresponds to the compactification scale $M_c$ of the theory.
At variance with 4D SUSY SU(5), in 5D threshold corrections 
typically fix $M_c$ around $5\times 10^{14}$ GeV, much smaller 
than the conventional grand unification scale. The proton decay
rate is eventually controlled by $M_c$ and by the gauge 
interactions of the light fermions with the heavy gauge bosons X.
These interactions are highly sensitive to the localization
of matter in the fifth dimension. Only fermions living
on a particular 4D slice of the space time have standard gauge
interactions to X. The assignment of fermions to such a slice
or to other specific locations of the extra space is a model-dependent
issue, tightly related to the description of fermion masses,
but not uniquely settled. 
More than a unique prediction, 5D SUSY SU(5) provides
several viable scenarios \cite{AFLV}, of great experimental interest.

The main motivation of the present note is to perform a similar 
analysis in the SO(10) case. Unlike in SU(5), for SO(10)
a minimal model in 5 dimension does not exist.
Early attempts have mainly dealt with 6 dimensions \cite{6D}, 
in order to exploit as much as possible the compactification mechanism
to break the SO(10) gauge symmetry. However,
to naturally achieve a DT splitting, it is sufficient
to work in a 5D setup \cite{MD,KR}, where SO(10) is broken by boundary 
conditions down to the Pati Salam (PS) group, at the extremum $y=\pi R/2$
of an interval $(0,\pi R/2)$ describing the fifth dimension.
The final breaking of PS down to the SM group can be realized
through an Higgs mechanism, also taking place at $y=\pi R/2$,
as indicated by a next-to-leading order RGE analysis.
This configuration can provide the basis for a minimal SO(10)
GUT in 5D, which we will complete and analyze hereafter.

First of all, after recalling how gauge symmetry breaking 
takes place, we will introduce matter fields and Yukawa couplings.
A correct description of fermion masses and mixing angles,
including those relevant to neutrino oscillation, is notoriously
a difficult task in SO(10) models where minimal Higgs content
gives rise to too rigid fermion mass relations.
In our proposal we will exploit the geometrical suppression
of Yukawa interactions associated to bulk matter fields,
by assigning first and second generation matter fields
to the bulk, with full dependence upon the extra coordinate $y$.
The heaviness of the third generation is guaranteed by locating
the corresponding multiplet at the PS brane. The choice of this
brane is made essentially unique by the requirement of breaking
the unwanted ``minimal'' SO(10) fermion mass relations.
To this purpose the third generation is described by several
irreducible PS representations, a 16 and a 10 from the SO(10)
point of view, giving to the Yukawa sector the desired flexibility.
To get rid of the additional matter degrees of freedom
such a system (16,10) has to be placed at the PS brane.
The overall picture reproduces, at the order-of-magnitude level,
all fermion masses and mixing angles and it is only compatible
with a semi-anarchical neutrino mass matrix, leading to a neutrino
spectrum of normal hierarchy type.
All the required relations are enforced by three suppression
parameters $\epsilon$, $\delta$ and $\epsilon_u$, the first two
being of geometrical origin. 
This completes a sort of minimal SO(10) 5D model, where the 
main features needed to estimate the proton lifetime are all
present.

It turns out that the prediction for proton decay is much more
constrained than in the corresponding SU(5) model.
Dimension 5 operators are still absent, and, due to the
specific localization properties of matter fields, necessary to
correctly reproduce fermion masses within a reasonably simple
framework, also minimal couplings of the super-heavy SO(10) gauge
bosons X and Y to matter are vanishing. Non minimal couplings 
are possible and here we provide our best estimate for them.
Unfortunately, the resulting dimension 6 operators describing
proton decay are depleted by the cut-off scale, too strong a
suppression to produce observable effects, even in future,
super-massive detectors. 

%
\section{SO(10) grand unification models in 5 dimensions}

We consider minimal supersymmetric SO(10) GUTs in 5 dimensions based on models constructed in \cite{MD, KR}.
The 5D space-time is factorized into a product of the ordinary 4D space-time M$_4$ and
of the orbifold $S^1/(Z_2\times Z_2')$, with coordinates $x^{\mu}$, ($\mu=0,1,2,3$) and $y=x^5$.
The fifth dimension lives on a circle $S^1$ of radius $R$ with the identification provided by the two reflections:
$ Z_2:y\rightarrow -y,$ and $Z_2':y'\rightarrow -y'$ with $y'\equiv y-\pi R/2$.
After the orbifolding, the fundamental region is the interval from $y=0$ to $y=\pi R/2$ with two inequivalent fixed points
at the two sides of the interval. The origin $y=0$ and $y=\pi R$ represent the same physical point and similarly for
$y=+\pi R/2$ and $y=-\pi R/2$. When speaking of the brane at $y=0$, we actually mean the two four-dimensional slices at
$y=0$ and $y=\pi R$, and similarly $y=\pi R/2$ stands for both $y=\pm\pi R/2$.

Generic bulk fields $\phi(x^{\mu},y)$ are classified by their orbifold parities $P$ and $P'$ defined by
$\phi(x^{\mu},y) \to \phi(x^{\mu},-y)=P\phi(x^{\mu},y)$ and $\phi(x^{\mu},y') \to \phi(x^{\mu},-y')=P'\phi(x^{\mu},y')$.
We denote by $\phi_{\pm \pm}$ the fields with $(P,P')=(\pm,\pm)$ with the following $y$-Fourier expansions:
\bea
  \phi_{++} (x^\mu, y) &=&
\sqrt{\frac1{2\pi R}}\phi^{(0)}_{++}(x^\mu)+
       \sqrt{\frac1{\pi R}}
      \sum_{n=1}^{\infty} \phi^{(2n)}_{++}(x^\mu) \cos{\frac{2n y} R}~~~,
\label{phi++exp}\nn\\
  \phi_{+-} (x^\mu, y) &=&
       \sqrt{\frac1{\pi R}}
      \sum_{n=0}^{\infty} \phi^{(2n+1)}_{+-}(x^\mu) \cos{\frac{(2n+1)y} R}~~~,
\label{phi+-exp}\nn\\
  \phi_{-+} (x^\mu, y) &=&
       \sqrt{\frac1{\pi R}}
      \sum_{n=0}^{\infty} \phi^{(2n+1)}_{-+}(x^\mu) \sin{\frac{(2n+1)y} R}~~~,
\label{phi-+exp}\nn\\
 \phi_{--} (x^\mu, y) &=&
       \sqrt{\frac1 {\pi R}}
      \sum_{n=0}^{\infty} \phi^{(2n+2)}_{--}(x^\mu) \sin{\frac{(2n+2)y} R}~~~.
\label{fourier}
\eea
where $n$ is a non negative integer. The Fourier component $\phi^{(n)}(x)$ of fields with opposite parities $(P,P')$
acquires a mass $(2n+1)/R$ upon compactification, while the component of fields with same parities acquires
a mass $(2n+2)/R$. Masses of the Kaluza-Klein modes are thus integer multiples
of the compactification scale $M_c=1/R$.
The gauge coupling unification depends crucially on the structure of the even and odd Kaluza-Klein (KK) towers.
Only $\phi_{++}$ has a massless component and only $\phi_{++}$ and $\phi_{+-}$ are non-vanishing on the
$y=0$ brane. The fields $\phi_{++}$ and $\phi_{-+}$ are non-vanishing on the $y=\pi R/2$ brane,
while $\phi_{--}$ vanishes on both branes.


The theory under investigation is invariant under N=1 SUSY in 5D,
which corresponds to N=2 in four dimensions, and under SO(10)
gauge symmetry. The gauge supermultiplet is in the adjoint
representation of SO(10) and can be arranged in an N=1 vector
supermultiplet $V$ and an N=1 chiral multiplet $\Phi$. We
introduce a bulk Higgs hypermultiplet in the fundamental
representation of SO(10), which consists of two N=1 chiral
multiplets $H_{10}$, $\hat H_{10}$ from a 4D point of
view.

Parities of the fields are assigned in such a way that
compactification reduces N=2 to N=1 SUSY and breaks SO(10) down to
the PS gauge group $\ps$. The $P$ and $P'$ assignments are given
in Table \ref{t1} \cite{MD, KR}. The breakdown of N=2 to N=1 is
quite simple and is achieved by the parity $P$. As illustrated in
Table \ref{t1}, $H_{10}$ and $V$ have even $Z_2$ parities, while
$\hat H_{10}$ and $\Phi$ have odd $Z_2$ parities and then vanish
on the brane $y=0$. The additional parity $P'$ respects the
surviving N=1 SUSY and can break the GUT gauge group. In fact, if
we denote the PS and the SO(10)/PS gauge bosons as  $V^+$ and
$V^-$ respectively, from the assignments of $Z_2'$ parities of
Table \ref{t1} for $V^+$ and $V^-$, it turns out that, on the
brane $y=\pi R/2$, only $V^+$ survives, with PS gauge symmetry.

The projection $Z_2'$ can furthermore split the Higgs chiral
multiplet $H_{10}$ ($\hat H_{10}$) in two chiral
multiplets{\footnote{ The PS gauge group is isomorphic to the product
$SO(6) \times SO(4)$.}}: $H_{10}=(H_6, H_4)$ ($\hat H_{10}=(\hat
H_{6},\hat{H}_{4}))$. $H_4$ contains scalar Higgs doublets $H^D_u$
and $H^D_d$ and $H_6$ contains the corresponding scalar triplets
$H^T_u$ and $H^T_d$. As an important consequence of the parity
assignments for the Higgs fields in Table \ref{t1}, only the Higgs
doublets and their superpartners are massless, while color
triplets and extra states acquire masses of order $1/R$, giving
rise to an automatic doublet-triplet splitting.
Notice that, had we used the $Z_2'$ projection to break SO(10)
down to SU(5)$\times$ U(1), we would have not achieved such an automatic
splitting. 

Gauge symmetry would allow a mass term for the $H_{10}$ on the brane $y=0$ or a mass term for the $H_4$ (and/or the $\hat H_{6}$)
on the brane $y=\pi R/2$ as pointed out in \cite{MD}, thus spoiling the lightness of the Higgs doublets achieved by compactification,
but such a term can be forbidden by explicitly requiring an additional
U(1)$_R$ symmetry \cite{hn1,hn2}.
Therefore, before the breaking of the residual N=1 SUSY, the mass spectrum is the one shown in Table 1.

\begin{table}[h]
{\begin{center}
\begin{tabular}{|c|c|c|}
\hline
& & \\
$(P,P')$ & field & mass\\
& & \\
\hline
& & \\
$(+,+)$ &  $V^+$, $H_4$ & $\frac{2n}{R}$\\
& & \\
\hline
& & \\
$(+,-)$ &  $V^-$, $H_6$ & $\frac{(2n+1)}{R}$ \\
& & \\
\hline
& & \\
$(-,+)$ &  $\Phi^-$, $\hat{H}_6$
& $\frac{(2n+1)}{R}$\\
& & \\
\hline
& & \\
$(-,-)$ &  $\Phi^+$, $\hat{H}_4$ & $\frac{(2n+2)}{R}$ \\
& & \\
\hline
\end{tabular}
\end{center}}
\caption{Parity assignment and masses ($n\ge 0$) of fields in the vector and Higgs supermultiplets. $V^+$ contains the PS gauge
bosons; $V^-$ contains the SO(10)/PS gauge bosons X and Y.}
\label{t1}
\end{table}


The further breaking of Pati-Salam gauge symmetry to the SM gauge group 
$G_{SM}$
cannot be obtained through an orbifold projection in five dimensions,
but it can be accomplished via brane-localized Higgs mechanism either on
the SO(10) brane \cite{MD} or on the PS brane \cite{KR}. 
In the first case, a pair of Higgs in the spinorial representation
$(\bf{16})+(\bf{\overline{16}})$ \footnote{The presence of both
those scalar multiplets is required in order to preserve the $N=1$
supersymmetry still present in the 4D theory after orbifolding.}
of SO(10) is introduced on the
brane $y=0$. In this way it is possible to break the gauge group
SO(10) down to SU(5), leaving \footnote{From the effective,
$4$-dimensional theory point of view.} the SM gauge group
unbroken: this happens since $G_{SM}$ is the intersection of SU(5)
and $\ps$. The second possibility is to reduce directly the PS
gauge group on the symmetry breaking brane $y= \pi R/2$. This can
be achieved by two Higgs multiplets $\Sigma$ and $\bar{\Sigma}$
in the representation $({\bf
\overline{4},1,2}) \oplus ({\bf 4,1,2})$ of $\ps$.

These two ways of realizing the Higgs mechanism on a brane will both
give the MSSM in the massless spectrum. However, different choices
of Higgs mechanism give completely different predictions for gauge
unification, as pointed out in Ref.~\cite{AL}. It has been shown
that only when the Higgses are localized on the PS brane with a
VEV near the cutoff scale, gauge coupling
unification is naturally preserved at the next to leading order. 
Therefore, to correctly achieve gauge coupling unification, we will only
concentrate on having $\Sigma$ and $\bar{\Sigma}$ localized on the
PS brane. In
order to obtain a 4-dimensional theory with the SM gauge symmetry
we give $\Sigma$ a huge, nonzero VEV, $u_{\Sigma}$, along the
direction of the SM singlet. The symmetry breaking
originated by the VEV of the brane field $\Sigma$ gives a
localized mass to the gauge fields
belonging to PS/SM \cite{AL, Nomura:2001mf}, without affecting the
spectrum of bulk Higgs fields. The effect of this high scale
localized mass term is to change the boundary condition for the
wave function, in such a way that the masses of the KK tower of
gauge bosons are shifted. Precisely, the KK mass spectrum of those
gauge bosons, subset of $V^+$ in Table~\ref{t1}, which belong to
PS/SM, is modified according to \begin{equation} \label{spectrum}
M_n=\frac{2n}{R} \rightarrow M_n \simeq \frac{2n+1}{R}
\left(1-\frac1a\right)~. \end{equation} where $ a  = \pi g_5^2
u_{\Sigma}^2 R/4 \gg 1$ and $n \geq 0 $. As a result, the
surviving gauge group is that of the SM and the massless spectrum
is exactly the MSSM one. By introducing the dimensionless parameter
\begin{equation}
x=\frac{g_5}{\sqrt{\pi\Lambda}} u_\Sigma~~~,
\label{defx}
\end{equation}
we can write
\begin{equation} 
\label{defa} 
a =\frac {x^2\pi^2}4 \frac {\Lambda} {M_c}~.
\end{equation}
In a strong coupling regime, naive dimensional analysis 
suggests $g_5^2 = 16\pi^3/\Lambda$ and
$\langle\Sigma\rangle = \Lambda/4 \pi$, which, in turn
gives $x=1$. The requirement of gauge
coupling unification gives a prediction on the compactification
scale $M_c$ which is important for our estimates of proton decay
rates. As pointed out in \cite{AL}, the gauge coupling
unification, and consequently $M_c$, depends strongly on the
values of $u_{\Sigma}$. For this reason, we will consider values
of $u_{\Sigma}$ more general than its naive dimensional value
$\Lambda/4 \pi$ and, in our estimates, we will allow $x \leq 1$.


\section{Fermion masses}

It is the matter content of the model under examination that will prove particularly interesting: indeed, it will be shown that a slight modification in the usual SO(10) setup allows to produce a phenomenologically interesting pattern for Yukawa matrices.

There are three possibilities to introduce quark and leptons in 5D orbifold constructions of SO(10). They can be described as N=1 SUSY chiral multiplets, localized on the two branes or introduced in the bulk as N=2 hypermultiplets. Either in the bulk or on the brane at $y = 0$, where the gauge group is unbroken, all matter fields should be introduced as complete SO(10) representations. Differently, on the symmetry breaking brane at $y =\pi R/2$ with the residual PS gauge group, matter fields should belong to $\ps$ representations. The choice between those various possible placements of the matter fields will be guided by the observed fermion mass hierarchies and mixings. This freedom is one of the new features of 5D orbifold GUTs. Six-dimensional proton decay operators arising from minimal or non-minimal couplings of X and Y gauge boson in SO(10) depend crucially on the localization of matter fields. On the other hand, the main flavor structure can be achieved without an ad hoc adjustment of the Yukawa couplings but only through geometrical suppression which naturally occurs in orbifold constructions.

\begin{table}[h]
{\begin{center}
\begin{tabular}{|c|c|c|}
\hline
& & \\
& {\tt bulk }  & {\tt brane $y=\pi R/2$} \\
& & \\
\hline
\hline
& & \\
{\tt Matter fields}& $\psi_i=(q_i,l_i,u^c_i,d^c_i,e^c_i,\nu^c_i) \qquad (i=1,2)$ & $\psi_3=(q_3,L) \oplus (u^c_3,D^c,e^c_3,\nu^c_3)$ \\
& & $\eta=(l_3,L^c) \oplus (d^c_3,D)$\\
& & \\
\hline
& & \\
{\tt Higgs sector}& $H_{10}=(h_u,h_d,H^3_u,H^3_d)$ & $\Sigma \oplus \overline{\Sigma} = ({\bf
4},{\bf 2},{\bf 1}) \oplus ({\bf \overline{4}},{\bf 1},{\bf 2})$ \\
& & \\
\hline
\end{tabular}
\end{center}}
\caption{Matter fields and their locations. Bulk fields should be doubled, to provide the
correct number of zero modes. In the table this doubling is understood. The matter
fields on the brane transform under the PS group as  $\psi_3 = ({\bf
4},{\bf 2},{\bf 1}) \oplus ({\bf \overline{4}},{\bf 1},{\bf 2})$ and $\eta = ({\bf
1},{\bf 2},{\bf 2})\oplus ({\bf 6},{\bf 1},{\bf 1})$. 
Capital letters denote heavy degrees of freedom.}
\label{localization}
\end{table}

We propose a simple and economical fermion mass pattern: the localization of matter
fields is that shown in Table~\ref{localization}. 
The bulk fields $\psi_i=(q_i,l_i,u^c_i,d^c_i,e^c_i,\nu^c_i)$ $(i=1,2)$ are the usual 16-plets of SO(10). 
Each of them accommodates a whole SM fermion generation plus a right-handed neutrino.
They describe the first and the second generations.
The third generation is fully localized on the PS brane and is contained in
four irreducible representations: 
$\psi_3 = (q_3,L) \oplus (u^c_3,D^c,e^c_3,\nu^c_3)$, transforming as $({\bf4},{\bf 2},{\bf 1}) \oplus ({\bf \overline{4}},{\bf 1},{\bf 2})$ and 
$\eta = (l_3,L^c) \oplus (d^c_3,D)$, transforming as $({\bf1},{\bf 2},{\bf 2})
\oplus({\bf 6},{\bf 1},{\bf 1})$ of $\ps$.
As we will see in the next section, the additional degrees of
freedom contained in $\psi_3$ and $\eta$, namely those described by the fields $L$, $L^c$, $D$ and $D^c$, get large masses and decouple from the low-energy physics. It is convenient to describe the third generation in this way to overcome the well-known difficulties related
to the fermion spectrum in minimal SO(10). We recall that, under the PS gauge group, the {\bf16} of
SO(10) decomposes as $ {\bf 16} = ({\bf 4},{\bf 2},{\bf 1}) \oplus ({\bf
\overline{4}},{\bf 1},{\bf 2})$ while ${\bf 10} = ({\bf 1},{\bf 2},{\bf 2})\oplus({\bf
6},{\bf 1},{\bf 1})$. Therefore
$\psi_3$ and $\eta$ fill exactly one {\bf16} and one {\bf10} representations of SO(10).

\subsection{Yukawa textures}

In this section, we will describe extensively how our model provides an explanation of
the fermion mass hierarchies and mixing angles by exploiting the geometrical suppressions
due to the different relative normalization of bulk and brane matter fields. We start our
analysis by writing down the most general superpotential
containing the leading terms in an expansion in inverse powers of $\sqrt{\Lambda}$.
In order to be consistent with the
orbifold construction, we have to extend the $U(1)_R$ symmetry to the matter sector. We
further impose an additional discrete $Z_3$ flavour symmetry on our superpotential under
which only $\eta$, $\Sigma$ and $\overline{\Sigma}$ are charged. The flavour symmetry breaking is implemented by a flavon $\theta$ singlet of SO(10), living at $y = \pi R/2$. The transformation properties of the various matter fields and  $\theta$ under $U(1)_R$ and $Z_3$ symmetries are shown in Table~\ref{symmetries}. 
\begin{table}[h!]
{\begin{center}
\begin{tabular}{|c|c|c|c|c|c|c|}
\hline
& & & & & &\\
Fields & $\psi_{1,2,3}$ & $\eta$  & $\Sigma$ & $\overline{\Sigma}$ & $H_{10}$ & $\theta$\\
& & & & & &\\
\hline
\hline
& & & & & &\\
$U(1)_R$ & 1 & 1 & 0 & 0 & 0 & 0 \\
& & & & & &\\
\hline
\hline
& & & & & &\\
$Z_3$ & $1$ & $\omega$ & $\omega^2$  & $\omega^2$ & $1$ & $\omega^2$  \\
& & & & & &\\
\hline
\end{tabular}
\end{center}}
\caption{$U(1)_R$ and $Z_3$ charges for matter fields and $\theta$.
The parameter $\omega = \text{exp}(i 2\pi/3)$ is the cubic root of unity.}
\label{symmetries}
\end{table}
The superpotential reads
\begin{equation}
W=kW_D+W_S+W_M+...
\end{equation}
\noindent
where $W_D$ is responsible for the Dirac mass terms of the light fermions, $W_M$ contains the
heavy Majorana neutrino masses and $W_S$ provides large mass terms for the extra degrees
of freedom. Dots denote sub-leading higher dimensional operators.
We allow for a generic overall dimensionless constant $k$ in $W_D$ that
characterizes the strength of the coupling between the matter fields and the $H_{10}$ multiplet.
With a schematic notation \footnote{The Latin indices $i, j, ... = 1, 2$ denote the first two generations.} we have:
\begin{eqnarray} 
W_D &=& \delta(y-\pi R/2) \left(\frac{1}{\Lambda^{1/2}}  \psi_3 \psi_3 H_{10} 
 +\frac{1}{\Lambda}  \psi_i \psi_3 H_{10} +\frac{1}{\Lambda^{3/2}} \psi_i \psi_j H_{10}\right)\nonumber\\
&+&\delta(y-\pi R/2) \left(
 \frac{1}{\Lambda^{3/2}}  \psi_3 \eta \overline{\Sigma}  H_{10}
+ \frac {1}{\Lambda^2}  \psi_i \eta \overline{\Sigma}  H_{10} \right) \label{WD}\\
W_S &=& \delta(y-\pi R/2) \left( \psi_3 \eta \Sigma + \frac{1}{\Lambda^{1/2}}  \psi_i \eta \Sigma  
+\frac{1}{\Lambda} \eta \eta  \Sigma \overline{\Sigma}
+\frac{1}{\Lambda} \eta \eta  \theta^2 \right) \nonumber \\
W_M &=& \delta(y-\pi R/2) \left( \frac{1}{\Lambda^2}\psi_3 \psi_3 \Sigma \Sigma \theta
+\frac{1}{\Lambda^{5/2}} \psi_i \psi_3 \Sigma \Sigma \theta
+\frac{1}{\Lambda^{3}} \psi_i \psi_j \Sigma \Sigma \theta + (\Sigma \rightarrow \overline{\Sigma}) \right)~.\nonumber
\end{eqnarray}
We assume that the only relevant Yukawa interactions are those present
at the brane $y=\pi R/2$ \footnote{This assumption, which is consistent 
within our general framework, can be made natural if the VEV of $H_{10}$ has a 
non trivial profile along the fifth dimension and is mainly concentrated
around $y=\pi R/2$.}.
Apart from $k$, we have omitted all other dimensionless coupling constants that we generically
assume to be of order one. Moreover, each term in the above expressions may stand
for several independent gauge invariant expressions. The factors of the cutoff scale $\Lambda$ in
the superpotential terms also take into account the fact that bulk fields $(\psi_{1,2}, H_{10})$ do have
different dimensions than brane fields $(\psi_3, \eta, \Sigma, \overline{\Sigma})$. The 
PS gauge symmetry is broken down to the SM one by the large VEV $u_{\Sigma}$ of the field $\Sigma$.
We anticipate that we expect $u_{\Sigma}$ not far from the cutoff scale:  $u_{\Sigma} \approx 0.05\Lambda$. Similarly, we assume that the $Z_3$ symmetry is broken at a high scale by the VEV of the flavon 
$\theta$ with $\langle \theta \rangle \lesssim u_{\Sigma}$: the consistency of this assumption with the observed fermion spectrum will be checked once we determine the neutrino mass matrix.
The $Z_3$ symmetry is introduced to suppress a possible dangerous mass term for $\eta$ \footnote{Such a term would indeed spoil the mechanism by which we obtain the lopsided structure for the down
quark/charged lepton mass matrices.}, such as $\eta\eta$. At the same time, $Z_3$ also suppresses the right-handed neutrino mass term in $W_M$ and controls the absolute mass scale of light neutrinos. As we will see in
Sec. \ref{numass&theta}, the flavon  needs to acquire a VEV $\langle \theta \rangle \approx 10^{15}~ \text{GeV}$ in order to reproduce the correct order of magnitude of the atmospheric mass square difference $\Delta m^2_{\text{atm}}$. Terms in $W_D$
will give rise to low energy Yukawa couplings for charged fermions and Dirac mass terms
for the neutrinos, after electroweak symmetry breaking. In addition to the usual SO(10)
invariant $\psi_i \psi_j H_{10}$, the field $\eta$ introduces new invariant terms involving $H_{10}$ 
which will be crucial for our construction of the fermion mass pattern. In order to determine which fields
become super-massive, we first consider the
effect of the large VEVs $u_{\Sigma}$ and $\langle \theta \rangle \lesssim u_{\Sigma}$ and, for the time being, we neglect the VEVs of $H_{10}$, which cause the final step of symmetry breaking. By focusing on the zero modes of the bulk fields, after integrating over the fifth coordinate $y$, we get
\begin{eqnarray}  \label{zeromodes}
W \approx W_S+W_M &=& L^c (L+\varepsilon_il_i+\varepsilon_ul_3)u_{\Sigma}
+ D (D^c+\varepsilon_id^c_i+\varepsilon_ud^c_3)u_{\Sigma}\nonumber\\
&+& (\nu^c_3\nu^c_3+\varepsilon_i\nu^c_i\nu^c_3+\varepsilon_i\varepsilon_j\nu^c_i\nu^c_j)\langle \theta \rangle \varepsilon^2_u~,
\end{eqnarray}
where $\varepsilon_u = u_{\Sigma}/ \Lambda$, and contributions of relative order $\langle \theta \rangle^2 / u^2_{\Sigma}$ have been neglected. The
constants $\varepsilon_i$ are suppression factors carried by the zero modes of bulk fields. If these
modes are constant in the extra dimensional coordinate $y$, the suppression factor is simply
 $\varepsilon \sim 1/ \sqrt{\pi\Lambda R}$. However, if the profile of the zero mode is not constant in $y$, the suppression factor can be different. For instance, if the bulk hypermultiplet has a kink mass $m$ with
the appropriate sign, the suppression factor becomes $\delta \sim e^{-\pi m R} \ll \varepsilon$. 
In order to produce the required hierarchy between the fermion masses of the first and second generation, we will exploit this freedom and we assume $\varepsilon_1 \approx \delta \ll \varepsilon_2 \approx \varepsilon$.

From Eq. (\ref{zeromodes}) we see that the all the right-handed neutrinos $\nu^c$ acquire large masses.
As we will discuss later on, these large Majorana masses combine with the light Dirac neutrino masses in
the see-saw mechanism. We also see that the fields $L^c, D$ and the
combinations $L+\varepsilon_il_i+\varepsilon_ul_3$ and $D^c+\varepsilon_id^c_i+\varepsilon_ud^c_3$ get a mass of order $u_{\Sigma}$ and decouple from
the low-energy theory. In the charged fermion sector, the light fields are 
$(q_i,q_3)$, $(u^c_i,u^c_3)$,$(e^c_i,e^c_3)$, $(l_i+\varepsilon_i L, l_3+\varepsilon_uL)$ and
$(d^c_i+\varepsilon_iD^c,d^c_3+\varepsilon_uD^c)$ {\footnote{We are not paying attention to the exact coefficients, but rather to the orders of magnitude.}}.
In the following these fields will
be approximated by taking the limit $\varepsilon_i,\varepsilon_u \rightarrow 0$ which will give results sufficiently accurate
for our purposes.

Now we turn to the properties of the yukawa textures, focusing on the $W_D$ term,
recalling that the electroweak Higgs doublets are contained in $H_{10}$. The (1, 2, 2) component
of $H_{10}$ will have to get nonzero VEVs in order to break the SU(2) subgroup of $G_{SM}$. We
denote the electroweak VEVs of the zero modes with $v_u$ and $v_d$ respectively. We only keep
the zero modes and we set to zero the heavy fields $L^c, D$, $L+\varepsilon_il_i+\varepsilon_ul_3$ and $D^c+\varepsilon_id^c_i+\varepsilon_ud^c_3$.
After integration over $y$, from Eq. (\ref{WD}) we get
\begin{eqnarray} \label{WDzero}
W_D &=&  [\varepsilon_i \varepsilon_j(u^c_iq^c_j+u^c_jq^c_i)+\varepsilon_iu^c_iq^c_3+\varepsilon_iu^c_3q^c_i+u^c_3q^c_3]\varepsilon v_u \nonumber\\
&+&  [\varepsilon_i \varepsilon_j(d^c_iq^c_j+d^c_jq^c_i)+\varepsilon_id^c_iq^c_3+\varepsilon_id^c_3q^c_i+d^c_3q^c_3]\varepsilon v_d \\
&+&  [\varepsilon_i \varepsilon_j(e^c_il^c_j+e^c_jl^c_i)+\varepsilon_ie^c_il^c_3+\varepsilon_ie^c_3l^c_i+e^c_3l^c_3]\varepsilon v_d \nonumber\\
&+&  [\varepsilon_i \varepsilon_j(\nu^c_il^c_j+\nu^c_jl^c_i)+\varepsilon_i\nu^c_il^c_3+\varepsilon_i\nu^c_3l^c_i+\nu^c_3l^c_3]\varepsilon v_d +  \cdots \nonumber 
\end{eqnarray}
where dots stand for subleading corrections. An interesting structure for the mass matrices
of fermions then emerges merely due to the localization of the various hypermultiplets
illustrated in Table \ref{localization}.

The mass matrix for the up sector comes from the first row of Eq.~(\ref{WDzero}) and, recalling that $\varepsilon_1 \approx \delta$ and $\varepsilon_2 \approx \varepsilon$, we get
\begin{equation}
\label{mu}
m_u = k\varepsilon \left(\begin{array}{ccc}
  \delta^2 & \delta \varepsilon & \delta \\
  \delta \varepsilon & \varepsilon^2 & \varepsilon \\
  \delta & \varepsilon & 1\\
\end{array} \right) v_u ~.
\end{equation}
To fit quark masses in the up sector we set
$\varepsilon \sim \lambda^2$, $\delta \sim \lambda^3 \div
\lambda^4$, $\lambda\approx 0.22$ being, as usual, the Cabibbo suppression
factor. We also notice that, to reproduce the overall mass scale and,
in particular, the top quark, we need to tune the overall strength $k$ of the $H_{10}$ coupling to
matter such that $k \geq \varepsilon^{-1}$. In other words we should assume that
the interactions of the multiplet $H_{10}$ with matter fields are in a strong coupling
regime. 

As for the charged lepton/down quark sector we get, with the same
localization for the $\psi_i \mbox{'s}$, the following texture,
coming from the third and fourth rows of Eq.~(\ref{WDzero}):
\begin{equation}
\label{ml}
m_l \approx m_d^T = k\varepsilon \left(\begin{array}{ccc}
  \delta^2 & \delta \varepsilon & \delta \varepsilon_{u}\\
  \delta \varepsilon & \varepsilon^2 & \varepsilon \varepsilon_{u}\\
  \delta  & \varepsilon  & \varepsilon_{u} \\
\end{array} \right) v_d
\end{equation}
Notice that $m_l$ and $m^T_d$ differ from $m_u$ because the third column carries an extra factor $\varepsilon_{u}$, which is the suppression factor for the $\eta$ hypermultiplet.
We get a good approximation of the experimental data provided we have
\begin{equation}
\varepsilon \sim \lambda^2 \sim \varepsilon_{u} ; \qquad \delta \sim \lambda^3.
\end{equation}

Finally, we have to deal with neutrino masses. Neutrino Dirac mass terms are given by
the second row of Eq.~(\ref{WDzero}) and in turn, taking into account the suppression factors,
this results in:
\begin{equation}
m_{\nu}^D =  k\varepsilon \left(\begin{array}{ccc}
  \delta^2 & \delta \varepsilon & \delta  \varepsilon_{u}\\
  \delta \varepsilon & \varepsilon^2 & \varepsilon \varepsilon_{u} \\
  \delta  & \varepsilon  & \varepsilon_{u} \\
\end{array}\right)v_u
\end{equation}
\noindent with the same relative suppressions of $m_l$. Heavy Majorana mass terms
for the $\nu^c$s arise from Eq.~(\ref{zeromodes}) and give rise to the mass matrix:
\begin{equation}
m_{\nu^c} = \varepsilon^2_u \langle \theta \rangle \left( \begin{array}{ccc}
  \delta^2 & \delta \varepsilon & \delta \\
  \delta \varepsilon & \varepsilon^2 & \varepsilon \\
  \delta & \varepsilon & 1 \\
\end{array} \right)
\end{equation}
where the same geometrical suppression of $m_u$ is present. By
usual see-saw mechanism, the light neutrino mass matrix reads
\begin{equation}
\label{mnu}
m_{\nu} = k^2 \frac{\varepsilon^2} {\varepsilon^2_u} 
\left( \begin{array}{ccc}
  \delta^2 & \delta \varepsilon & \delta\varepsilon_u \\
  \delta \varepsilon & \varepsilon^2 & \varepsilon \varepsilon_u\\
  \delta \varepsilon_u& \varepsilon \varepsilon_u & \varepsilon_u^2 \\
\end{array} \right) \frac{v^2_u}{\langle \theta \rangle}~.
\end{equation}

\subsection{Neutrino masses and the VEV of $\theta$}
\label{numass&theta}

Our model predicts a specific pattern
for the neutrino mass matrix, also known as ``semi-anarchy'' \cite{af2}.
This structure can be consistent with experimental data provided
we assume that neutrino masses are hierarchical and that the solar
mixing angle is somewhat enhanced. Explicitly, since we have required
that $\varepsilon \sim \varepsilon_u$, we have for the neutrino
mass matrix \footnote{Note that, having assumed $\varepsilon \sim
\lambda^2$ and $\delta \sim \lambda^3 \div \lambda^4$,
automatically we get $\delta / \varepsilon$ to be small, and
therefore our $m_{\nu}$ does indeed reproduce the semi-anarchy
structure.}
\begin{equation}
\label{mnusimpl}
 m_{\nu} =
 \left( \begin{array}{ccc}
  \delta^2 / \varepsilon^2 & \delta / \varepsilon & \delta / \varepsilon \\
  \delta/ \varepsilon & 1 & 1\\
  \delta / \varepsilon & 1 & 1 \\
\end{array} \right) \frac{v_{u}^2} {\langle \theta \rangle}
\end{equation}

\noindent 
and, apart from the overall mass scale $v_{u}^2
/\langle\theta\rangle$, the determinant of the 23 block in
$m_\nu$, which is generically of order one, should be tuned around
$m_2/m_3\approx\sqrt{\Delta
m^2_{sol}/ \Delta m^2_{atm}}\approx 0.1\div 0.2$.

The mass matrix in eq. (\ref{mnusimpl}) predicts a neutrino spectrum of the normal hierarchy type.
Thus the overall scale is determined by the atmospheric squared mass difference: 
\begin{equation}
\frac {{v_u}^2} {\langle \theta \rangle}
\sim \sqrt{\Delta m^2_{atm}} \sim 5 \cdot 10^{-2} \text{eV}
\end{equation}
\noindent 
and, if we take $v_u \sim 100~
 \text{GeV}$, we get
$\langle\theta \rangle \sim 10^{15}~ \text{GeV}$.
As we will see in Sec. \ref{gcu} by discussing gauge coupling unification,
the central value of the cut-off $\Lambda$ is around of $10^{17} ~\text{GeV}$
and $\langle\theta \rangle$ is 
about two orders of magnitude below such a scale.
Moreover, as we have seen above by discussing the fermion textures, 
we need $\langle \Sigma \rangle$ about a factor 20 below $\Lambda$. 
We conclude that $\langle \theta \rangle \lesssim \langle \Sigma \rangle$.

\subsection{Fermion mixings}
The quark mixing matrix $V_{CKM}$ and the lepton one $U_{PMNS}$
are given by:
\begin{equation}
V_{CKM}=L_u^\dagger L_d~,~~~~~~~U_{PMNS}=L_e^\dagger L_\nu~,
\end{equation}
where the unitary rotations $L_u,L_d, L_e$ map left-handed charged fermions from
the interaction  basis into the mass eigenstate basis:
\begin{equation}
\begin{array}{ccc}
u\to L_u u & d\to L_d d & e\to L_e e~,
\end{array}
\label{mapL}
\end{equation}
and $L_{\nu}$ diagonalizes the symmetric light neutrino mass
matrix $m_{\nu}~$. In analogy to what happens with the L matrices,
right-handed charged fermions are rotated to mass eigenstates by R
matrices:
\begin{equation}
\begin{array}{cccc}
u^c\to R_u^\dagger u^c & d^c\to R_d^\dagger d^c &
e^c\to R_e^\dagger e^c
\end{array}~.
\label{mapR}
\end{equation}
Right-handed rotation matrixes R have no observable consequences
in oscillation experiments, but, in general, they are important in
estimating proton decay rates in orbifold construction
\cite{AFLV}. $L$ and $R$ matrices can be estimated from
Eqs.~(\ref{mu},~\ref{ml},~\ref{mnu}). Assuming that $\varepsilon_u
\sim \varepsilon \sim \lambda^2$ and $\delta$ varying in the range
$\lambda^3\div \lambda^4$, we find:
\begin{eqnarray}
L_u\equiv R_u\approx L_d\approx R_e &\approx&
\left(
\begin{array}{ccc}
1 & \lambda & \lambda^3 \\
\lambda & 1 & \lambda^2 \\
\lambda^3 & \lambda^2 & 1
\end{array}
\right)
\div
\left(
\begin{array}{ccc}
1 & \lambda^2 & \lambda^4 \\
\lambda^2 & 1 & \lambda^2 \\
\lambda^4 & \lambda^2 & 1
\end{array}
\right)~,
\label{luld} \\
L_e\approx R_d &\approx& \left(
\begin{array}{ccc}
1 & \lambda & \lambda\\
\lambda & 1 & 1\\
\lambda & 1 & 1
\end{array}
\right)
\div
\left(
\begin{array}{ccc}
1 & \lambda^2 & \lambda^2\\
\lambda^2 & 1 & 1\\
\lambda^2 & 1 & 1
\end{array}
\right)~,
\label{le}\\
L_\nu
&\approx& \left(
\begin{array}{ccc}
1 & 1 & \lambda\\
1 & 1 & 1\\
\lambda & 1 & 1
\end{array}
\right)
\div
\left(
\begin{array}{ccc}
1 & \lambda & \lambda^2\\
\lambda & 1 & 1\\
\lambda^2 & 1 & 1
\end{array}
\right)~.
\end{eqnarray}
These expressions for $L_u$, $L_d$, $L_e$, $L_{\nu}$ allow us
to easily estimate the fermion mixing matrices:
\begin{eqnarray}
\text{for~} \delta &\approx& \lambda^4 \qquad V_{CKM} \approx \left(
\begin{array}{ccc}
1 & \lambda^2 & \lambda^4 \\
\lambda^2 & 1 & \lambda^2 \\
\lambda^4 & \lambda^2 & 1
\end{array}
\right)
\qquad
U_{PMNS} \approx
\left(\begin{array}{ccc}
1 & \lambda & \lambda^2\\
\lambda & 1 & 1\\
\lambda & 1 & 1
\end{array}
\right)~,
\label{case1}\\
\text{for~} \delta &\approx& \lambda^3 \qquad V_{CKM} \approx
\left(
\begin{array}{ccc}
1 & \lambda & \lambda^3 \\
\lambda & 1 & \lambda^2 \\
\lambda^3 & \lambda^2 & 1
\end{array}
\right)
\qquad
U_{PMNS}\approx \left(
\begin{array}{ccc}
1 & 1 & \lambda\\
1 & 1 & 1\\
1 & 1 & 1
\end{array}
\right)~.
\label{case2}
\end{eqnarray}

In our estimate of the proton lifetime we will consider both cases 
$\delta\approx \lambda^4$ and $\delta\approx \lambda^3$, though
the final results are not too much sensitive to this variation.
 

\section{Gauge coupling unification}
\label{gcu}
A next to leading analysis of gauge coupling unification of this
model has been discussed in ref. \cite{AL}. Here we will 
summarize the main points and the results.
The low-energy coupling constants
$\alpha_i(m_Z)$ $(i=1,2,3)$ in the $\overline{MS}$ scheme are
related to the unification scale $\Lambda$, the common value
\footnote{Strictly speaking, the gauge coupling constants never unify and
$g_U$ represents only a mean value. Exact gauge coupling unification
at the cut-off scale is spoiled by SO(10) breaking effects
occurring on the SO(10) violating brane and 
included in the present analysis. Under certain conditions
these effects are small and do not spoil the predictability 
of the model.}
$\alpha_U=g_U^2/(4\pi)$ at $\Lambda$ 
and the compactification
scale $M_c$ by the renormalization group equations (RGE):
\begin{equation}
\frac{1}{\alpha_i(m_Z)}=\frac{1}{\alpha_U} +\frac{b_i}{2\pi}\log
\left(\frac{\Lambda}{m_Z}\right) + \delta^{NL}_i~. \label{rge}
\end{equation}
\noindent
Here $b_i$ are the coefficient of the SUSY $\beta$ functions at
one-loop, $(b_1,b_2,b_3)=(33/5,1,-3)$, for 3 generations and 2
light Higgs SU(2) doublets. We recall that $g_1$ is related to the
hypercharge coupling constant $g_Y$ by $g_1=\sqrt{5/3}~ g_Y$.
Since gauge coupling unification does not depend on the universal
contribution to the $\beta$ functions, we will 
subtract a universal constant from $b_i$ and we define
\footnote{We are indeed just taking into
account the differential running of the coupling constants}:
$b_1=0$, so $b_i=(0,-28/5,-48/5)$. In eq. (\ref{rge}),
$\delta^{NL}_i$ stand for non-leading contributions:
\begin{equation}
\delta^{NL}_i=\delta^{(2)}_i+\delta^{(l)}_i+\delta^{(h)}_i+\delta^{(b)}_i~~~,
\label{delta} \end{equation}
where $\delta^{(2)}_i$ represent two-loop running
effects, coming from the gauge sector \cite{twoloop},
$\delta^{(l)}_i$ are light threshold corrections at the SUSY
breaking scale \cite{susyth}, $\delta^{(h)}_i$ are heavy threshold
corrections at the compactification scale $M_c$ and finally
$\delta_i^{(b)}$ are unknown SO(10)-violating contributions originated 
by kinetic terms for the gauge bosons of $\ps$ on the brane at 
$y=\pi R/2$ \cite{AL}.

\begin{figure}[h!]
\begin{center}
\includegraphics[width=7.9cm]{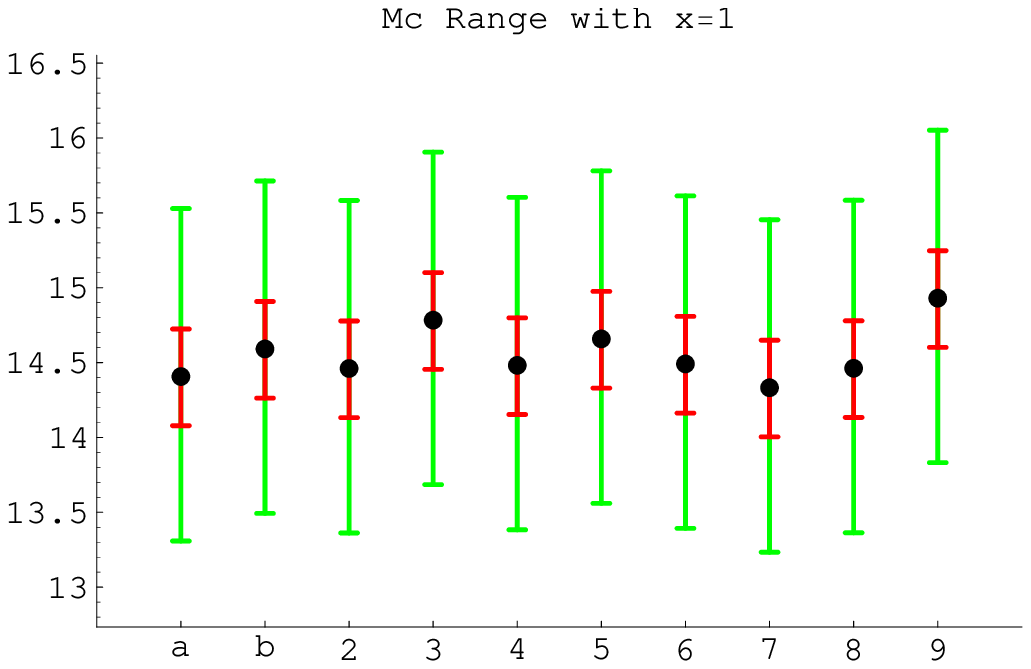}
\includegraphics[width=7.9cm]{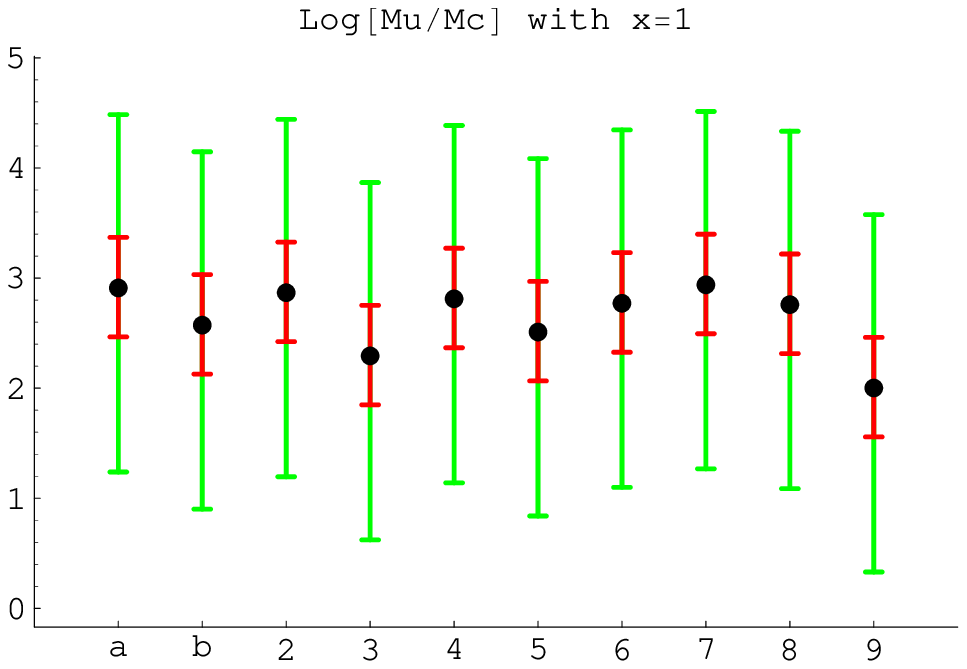}
\end{center}
\caption{Compactification scale $\log_{10}(M_c)$ and $\log_{10}(\Lambda/M_c)$ 
($M_u\equiv \Lambda$) versus SUSY spectrum, for $x=1$. The shorter error bars represent the
parametric error dominated by the experimental uncertainty on
$\alpha_3(m_Z)$, the wider bars include the dominant source of
error, the SO(10)-breaking brane terms $\delta_i^{(b)}\in
\left[-1/2\pi,+1/2\pi\right]$.} \label{rabi1}
\end{figure}

\begin{figure}[!h]
\begin{center}
\includegraphics[width=7.9cm]{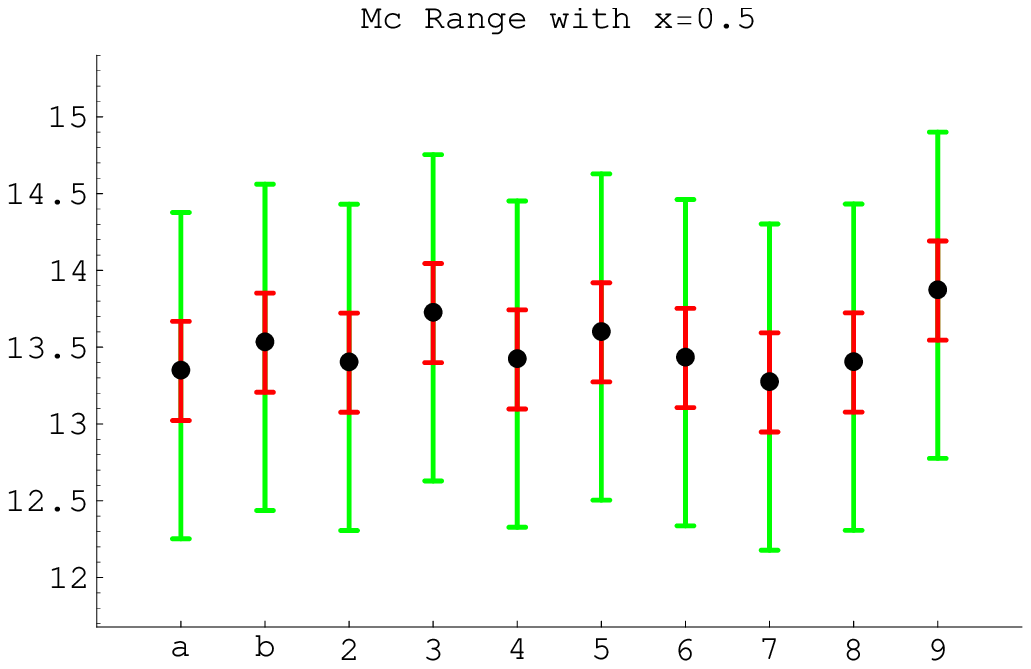}
\includegraphics[width=7.9cm]{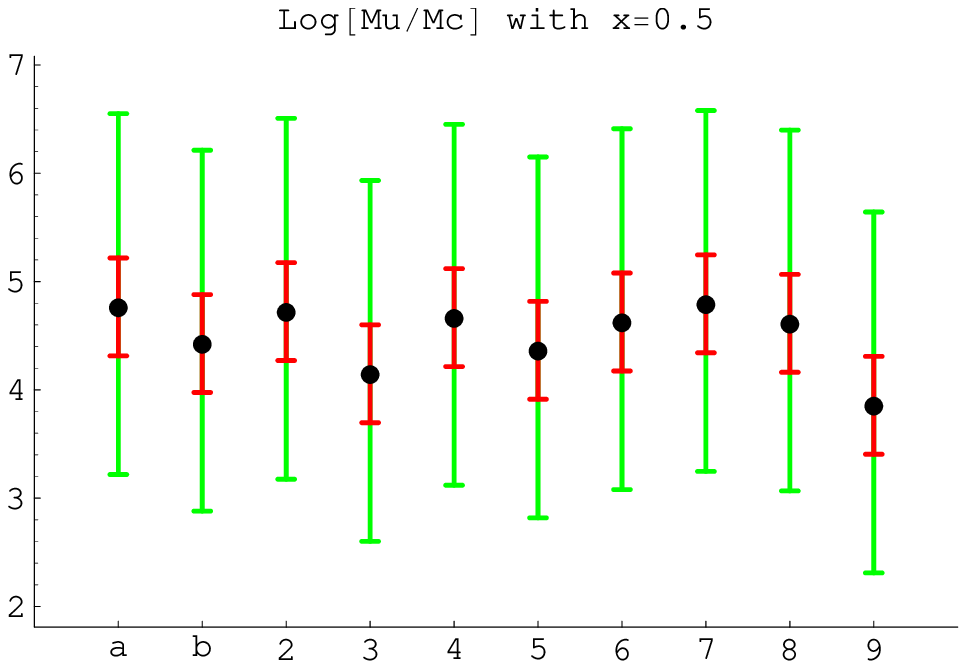}
\end{center}
\caption{Compactification scale $\log_{10}(M_c)$ and $\log_{10}(\Lambda/M_c)$
($M_u\equiv \Lambda$)
versus SUSY spectrum for $x=1/2$. Error
bars as in Fig.~{\ref{rabi1}}.} \label{rabi0.5}
\end{figure}

It is well-known that
the two-loop contributions and the threshold effects due to the 
light particles tend to raise the good leading-order prediction
of the strong coupling constant, $\alpha_3^{LO}(m_Z)\approx 0.118$:
\begin{equation}
\alpha_3(m_Z)=\alpha_3^{LO}(m_Z)[1-\alpha_3^{LO}(m_Z)\delta_s]~~~,
\end{equation}
where $\delta_s$ is a combination of $\delta_i^{NL}$ ($i=1,2,3$) \cite{AFLV}.
Two-loops and light thresholds give approximately
\begin{equation}
\delta_s^{(2)}\approx -0.82
~~~,~~~~~~~~
\delta_s^{(l)}\approx -0.50+\frac{19}{28\pi}\log\frac{m_{SUSY}}{m_Z}~~~,
\end{equation}
where $m_{SUSY}$ denotes the average mass of the light superpartners.
In the absence of additional effects we have $\alpha_3(m_Z)\approx 0.130$,
too large to be compatible with the present experimental value.
In the present model there are two other contributions. One comes from
SO(10) violating kinetic terms on the PS brane. They are due to unknown
ultraviolet physics and are expected to produce an effect of order
$\delta_i^{(b)}\approx \pm 1/2\pi$. This effect enhances the theoretical error 
on $\alpha_3(m_Z)$ but does not change its central value.
The second one comes from the thresholds associated to the heavy
particles.
The next-to-leading renormalization group evolution of the
coupling constants receives heavy thresholds contributions
$\delta^{(h)}_i$ from KK modes at the compactification scale
$M_c$. These can be computed in a leading
logarithmic approximation, including all states in the
Kaluza-Klein towers of gauge bosons and Higgs fields with masses
smaller than the cut-off scale $\Lambda$. The heavy threshold
contributions are given by \cite{AL}: \begin{equation}
\delta^{(h)}_i \approx
\frac{\alpha_i}{2\pi} \sum_{n=0}^N \log\frac{(2n+2)}{(2n+1)}+
\frac{\beta_i}{2\pi}\frac{(2N+2)}{a}~, \label{ht} \end{equation}
with $(\alpha_1,\alpha_2,\alpha_3)=(0,16/5,36/5)$, 
$(\beta_1,\beta_2,\beta_3)=(0,14/5,9/5)$ and $N$
defined by $(2 N+2)\approx \Lambda / M_c$.
The parameter $a$ is given in eq. (\ref{defa}), where we take $x\le 1$.
We recall that $x$ accounts for the dependence of the gauge coupling unification on $u_{\Sigma}$. If $x=1$ the theory is strongly coupled at 
the cut-off scale.
For large $N$, that is for $\Lambda R\gg 1$
\begin{equation}
\label{approx} \sum_{n=0}^N
\log\frac{(2n+2)}{(2n+1)} \approx
\frac{1}{2}\log(N+1)+\frac{1}{2}\log\pi \approx
+\frac{1}{2}\log\frac{\Lambda}{M_c}+\frac{1}{2}\log\frac{\pi}{2}~.
\end{equation}
In this limit
the heavy thresholds (\ref{ht}) become:
\begin{equation} \delta^{(h)}_i \approx
\frac{\alpha_i}{4\pi}\log\frac{\Lambda}{M_c}+\frac{\beta_i}{2\pi}\frac4{x^2
\pi^2}+... ~~~, \label{htaapprox}
\end{equation} where dots stand for
universal contributions. 
The shift in the strong coupling constant is given by:
\begin{equation}
\delta_s^{(h)}=\frac{3}{7 \pi}\log\frac{\Lambda}{M_c}-\frac{6}{\pi^3 x^2}~~~.
\end{equation}
This contribution can bring $\alpha_3(m_Z)$ back into the experimentally
allowed interval, provided $\Lambda/M_c$ is sufficiently large
and $x$ not too small. This constrains the compactification
scale $M_c$.
Numerical evaluation of $M_c$ can be
easily performed by running the next to leading RGE (\ref{rge})
using experimental data for the low energy values of the coupling
constants: \begin{eqnarray}
\alpha_{em}^{-1}(m_Z)&=&127.906\pm 0.019\nn\\
\sin^2\theta_W(m_Z)&=&0.2312\pm 0.0002\nn\\
\alpha_3(m_Z)&=&0.1187\pm 0.0020~ \label{inputrge}
\end{eqnarray}
Detailed numerical results {\footnote{For more details, see \cite{AL}.}} for $x=1$ and $x=1/2$ are plotted
in Fig.
(\ref{rabi1},~\ref{rabi0.5}). As in \cite{AFLV}, we parametrize our ignorance 
about the SUSY breaking mechanism, assuming a variety of supersymmetric particle spectra, 
and for each of them we evaluate $\log_{10}(M_c)$ and 
$\log_{10}(\Lambda/M_c)$.
We have adopted the so-called Snowmass Points and Slopes (SPS), derived from Ref.~\cite{SPS},
which are a set of benchmark points and parameter lines in the MSSM parameter space 
corresponding to different scenarios. 
The ten different spectra are listed in Table 4 of Ref.~\cite{AFLV}.
The compactification scale is very sensitive to  $\langle\Sigma\rangle$, that is to $x$, and for
$x=1$ its central value is $M_c \approx 3 \times 10^{14}
~\text{GeV} $ which is approximately two orders of magnitude
smaller than the 4D unification scale $M_U$. Values of
$M_c$ would be furthermore lowered for values of $x$ smaller than 1,
which could be potentially very dangerous for proton 
lifetime. Indeed, as we will see in section \ref{pd}, because of
our choice of matter field localization X and Y gauge bosons never
mediate directly baryon-violating processes. Moreover, non minimal
six-dimensional operators will be found to be heavily suppressed.

\section{Proton lifetime}\label{pd}

In our model proton decay is dominated by heavy gauge boson exchange.
Indeed, dimension 5 operators arising through coloured higgsino
exchange are forbidden by the 
U(1)$_R$ R-symmetry of the 5D theory, only broken around the
electroweak scale. We assume that, as in
the MSSM, the R-parity subgroup of the U(1)$_R$ symmetry
remains an exact symmetry of the low-energy theory, thus
prohibiting renormalizable baryon-violating operators as well. 
Therefore proton decay mainly proceeds through the exchange of the 
gauge bosons X and Y of the vector 
supermultiplet $V^-$ belonging to SO(10)$/$PS
\footnote{PS gauge bosons do not induce dimension 6 baryon-violating 
operators. Through a mixing with the X and Y gauge bosons they
can give rise to $\Delta B=-\Delta L$ dimension 7 operators,
completely negligible in the present model.}. 
Due to momentum conservation along the fifth dimension,
preserved by bulk interaction terms, X and Y gauge
bosons cannot couple to two zero modes of bulk hypermultiplets
through minimal gauge interactions.
They cannot minimally couple to zero modes on the PS brane either,
since X and Y vanish at $y=\pi R/2$.
The only zero modes that may have a minimal coupling to X and Y are
those described by matter fields on the SO(10) brane.
In our model matter fields are localized in the
bulk or on the PS brane, see Table~\ref{localization}, and consequently
dimension 6 baryon-violating minimal interactions are absent by
construction.
We stress that this conclusion is strictly related to our discussion 
of gauge coupling unification. As discussed in ref. \cite{AL}, 
a correct next-to leading order gauge coupling unification
requires the Higgs mechanism to take place on the PS brane.
As a consequence, to properly break the unwanted SO(10) fermion
mass relations, the simplest possibility is to accommodate
the third generation on the PS brane and the other generations 
in the bulk, thus preventing proton decay via minimal coupling.
The only possibility we are left with is to introduce non-minimal
interactions of X, Y gauge bosons with matter fields. 
Considering the lowest possible dimension, 
there are two type of operators that violate the baryon number.
\begin{itemize}

\item Type I:

Proton decay can arise from a derivative interaction 
localized on the PS brane:
\begin{equation}
{\cal L}_I =
\frac{c}{\Lambda} \delta(y-\frac{\pi R}2) \int d^2 \theta d^2 \bar \theta \
(\varphi)^{\dagger} \ (\nabla_5 e^{2 V}) \ \varphi'  + h.c.
\label{operator1}
\end{equation}
where $\nabla_5 =\partial_5 + \Phi$ and $(\varphi,\varphi')$
stands for any combination of $\psi_3$ and $\eta$.

\item Type II

Another contribution to proton decay can be originated by
non-diagonal kinetic terms between two members of the doubling
($\psi_i$, $\psi_i'$), $i=1,2$ of bulk fields. These operators are
localized on the SO(10) symmetric brane:
\begin{equation}
{\cal L} _{II} =
\frac{c'}{\Lambda} \delta(y) \int d^2 \theta d^2 \bar \theta \
(\psi_i)^{\dagger} (e^{2 V}) \ \psi'_i  + h.c. \label{operator2}
\end{equation}

\end{itemize}

The unknown constants $c$ and $c'$ are expected to be of order one and 
are free parameters in the effective theory. Therefore
proton lifetime cannot be
calculated accurately and here we can only give a crude,
order-of-magnitude, estimate, based on the leading operators
(\ref{operator1}, \ref{operator2}) of
our model. 
After integrating out the super heavy gauge bosons X, Y
($M_{X,Y}=2 M_c /\pi$), from Type I operators 
we obtain the four-fermion lagrangian:
\begin{equation} 
{\cal L}_p \sim \frac{c^2 g_U^2}{2
M_X^2} \frac1{\Lambda R} \left(a \overline{u^c_3} \gamma^\mu q_3
\cdot \overline{e^c_3} \gamma_\mu q_3 +b \overline{u^c_3}
\gamma^\mu q_3 \cdot \overline{d^c_3} \gamma_\mu l_3 \right)~~~,
\label{type1}
\end{equation}
where $a,b$ are dimensionless coefficient of order one and 
only the third generation is present \cite{hebecker1}. 
From Type II operators we get
\begin{equation} 
{\cal L}_p \sim
-\frac{(c')^2 g_U^2}{2 M_X^2} \frac1{(\Lambda R)^2} \left(
\overline{u^c_i} \gamma^\mu q_i \cdot \overline{e^c_i} \gamma_\mu
q_i -2 \overline{u^c_i} \gamma^\mu q_i \cdot \overline{d^c_i}
\gamma_\mu l_i \right)~~~,
\label{type2}
\end{equation} 
where $i=1,2~$ \cite{hn2}.

Despite having the same dimension, the operators 
(\ref{type1}) and (\ref{type2}) have a different cut-off dependence.
The derivative interaction (\ref{operator1}) produces a
relative enhancement of order $\Lambda R$. This can be easily
understood by working in momentum space where the $1/n^2$
factor coming from the exchange of the $n$th KK mode is compensated
by an $n^2$ factor coming from the interaction vertices.
We are left with an unsuppressed sum over the KK modes,
that can be regularized by cutting the upper limit of the sum at
the KK mode whose mass exceeds the cut-off $\Lambda$. 
The sum gives approximately $\Lambda R$, the number
of KK modes below the cut-off. 
However, the parametric enhancement of the operator (\ref{type1})
is not sufficient to overcome the huge suppression factor 
coming from the flavor mixing angles required to rotate third generation 
fields into fields relevant to proton decay. 
A detailed numerical analysis shows that the contribution from 
Type I operators is very suppressed
and proton decay from this channel is beyond the possibilities 
of the next generation of experiments.
We recall that the present experimental bound on the proton lifetime
in the decay channel $e^+ \pi^0$ is $\tau(p\to \pi^0 e^+)> 5.4\times 
10^{33}$ yr (90\% C.L.), while the aimed for sensitivity of future
experiments in the same channel is close to $10^{35}$ yr.

\begin{figure}
\begin{center}
\includegraphics[width=7.9cm]{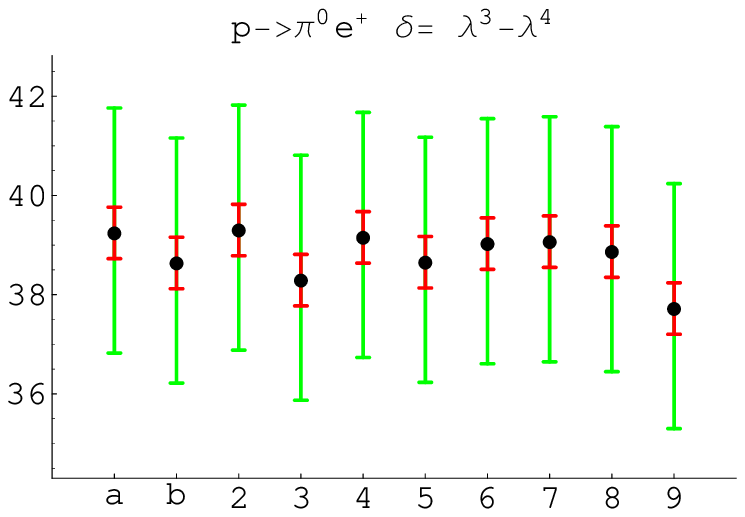}
\includegraphics[width=7.9cm]{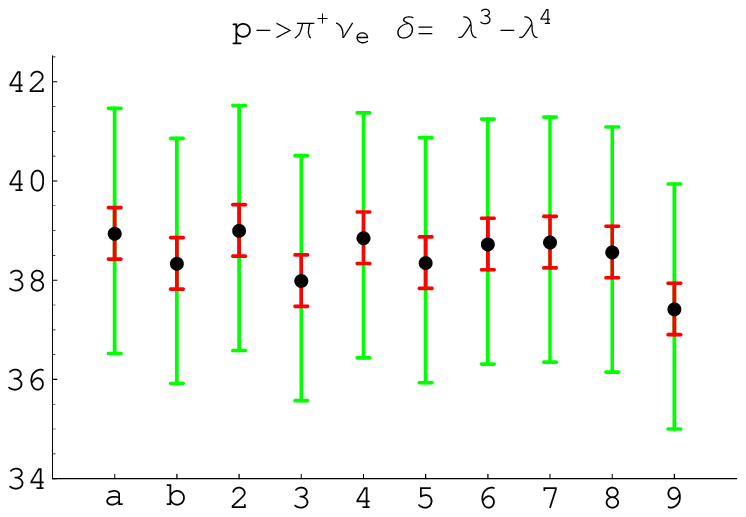}
\end{center}
\caption{Proton lifetime versus SUSY spectrum for the dominant channels from Type II operators (we display the $\log_{10}$ of the lifetime in years). The smallest error bar is due to the experimental error from $\alpha_3$, while the biggest one derives from the theoretical uncertainties concerning the SO(10)- breaking brane terms $\delta_i^{(b)}\in \left[-1/2\pi,+1/2\pi\right]$. (For more details look at \cite{AFLV}.)}
\label{TipoII}
\end{figure}

Type II operators are less suppressed but 
their contribution 
\footnote{Crossed contributions, obtained via gauge boson exchange
between an interaction of type I and an interaction of type II
are also negligible.} to proton decay is marginal.
Qualitatively this can be understood from the fact that
the mass scale suppressing dimension 6 operators in (\ref{type2}) 
is the combination $M_X \Lambda R\approx \Lambda$, which is
much larger than the compactification scale $M_X$.
The results for the dominant decay channels in this final case are 
shown in Fig. \ref{TipoII}.
We notice that the best values for the proton lifetime in the two 
channels displayed in Fig.(\ref{TipoII}) are not so different 
from the analogous values found in the SU(5) model analyzed in
ref. \cite{AFLV}.
The main difference deals with the theoretical error bar that 
in the present case covers about 4 orders of magnitude, 
while in the SU(5) case the uncertainty due to the unknown 
SU(5)-breaking brane contribution is much larger.
The reduction in the theoretical uncertainty is here due to
the fact that the amplitude is essentially controlled by the
cut-off scale $\Lambda$, rather than the compactification scale 
as in SU(5), and $\Lambda$ is less sensitive than $M_X$ to
the unknown SO(10) violating brane terms.
Unfortunately the net result is that
the possibility of seeing proton decay in this model 
and for these particular decay channels is even more disfavoured than
in the SU(5) model.

\section{Discussion}

We have searched for a simple and semi-realistic realization
of the SO(10) grand unified symmetry in the context of a 5D theory.  
To take advantage of the solution to the DT splitting problem
offered by models with extra dimensions, it is sufficient
to consider a single extra dimension and to reduce
SO(10) down to the PS group by compactification on $S^1/Z_2$.
The further breaking of the gauge symmetry down to the SM gauge group 
can be achieved through the ordinary Higgs mechanism taking place 
on the PS brane. In this note we have discussed in detail
the fermion spectrum of the model with the hope of reproducing
all the known features, at least at the level of orders of
magnitude.
Achieving a correct picture of fermion masses and mixing angles
in an SO(10) GUT is not an easy task, not even at the level of
a crude description. At variance with SU(5), where
the theory with minimal field content can already accommodate
a good first order approximation of the observed fermion spectrum,
we cannot speak of a ``minimal'' SO(10) model.
The reason is that in SO(10) with a minimal field content 
involving only the couplings $\psi_i \psi_j H_{10}$, 
the relation $M_u\propto M_d$ is completely wrong. Therefore,
we need not only some mechanism to naturally produce 
hierarchies between Yukawa parameters, but also a certain
degree of non-minimality.

In a 4D theory, hierarchies can be easily obtained by 
exploiting abelian flavour symmetries of the Froggatt-Nielsen 
(FN) type \cite{FN}, and a variety of textures for fermion 
mass matrices in SU(5) GUTs have been successfully 
constructed \cite{su5FN} along these lines. 
Gauge theories formulated in more than four space-time dimensions 
offer alternative possibilities. Hierarchies between Yukawa couplings 
can be generated by the geometrical properties of the extra space,
where matter fields may be localized in a variety of possible ways.
In the case of a single extra dimension 
represented by an interval of length $L$, zero modes of bulk fields
and brane fields enter Yukawa couplings with a relative normalization
given by a factor $1/\sqrt{\Lambda L}$. Already in this simple case
the observed hierarchy between fermion masses can be related to 
the hierarchy between the cut-off $\Lambda$ and the compactification 
scale $1/L$.
Various attempts have been made in this direction based on SU(5) in 5D 
\cite{AFLV, hebecker3}. Both in the case of abelian 4D flavour 
symmetries and in higher dimensional GUTs, it is not easy to accommodate
realistic fermion masses if all matter fields of the 
same family belong to a single GUT representation, as 
it happens in SO(10).
All fermions of a given generation tend to have similar Yukawa couplings
and it is difficult to generate different
hierarchical patterns in the different sectors.

To proceed towards a realistic model, some lessons can be drawn from the 4D 
case.
One way to reproduce phenomenologically viable mass patterns 
of fermions in SO(10) is to depart from the ``minimal'' SO(10)
Yukawa interactions by including higher representations of Higgs fields,
which can correspond to new elementary degrees of freedom
or to composite fields. 
In the ordinary 4D GUTs, this is a very popular approach and various 
realizations of Yukawa superpotentials, renormalizable or not, 
have been considered in the literature. 
In this direction, we find two kinds of realistic 4D SO(10) 
models. The first one is based on a relatively ``compact'' Higgs sector 
(for instance \{{\bf 16}, $\overline{\text{\bf 16}}$, {\bf 45}, {\bf 10}\}) and nonrenormalizable 
superpotentials \cite{so10type1}. In the second type of models 
the Higgs sector includes $H_{\overline{\text{126}}}$,
a 126 SO(10) representation, and the minimal Yukawa interactions 
are modified by adding a new renormalizable contribution of the type 
$\psi_i \psi_j H_{\overline{\text{126}}}$ \cite{so10type2}.
In both cases, an ad hoc pattern of Yukawa couplings is introduced 
in order to fit the experimental data. It should be said 
that in all SO(10) GUTs with extended Higgs sector the Higgs 
superpotential  is quite complicated and obtaining the desired 
gauge symmetry breaking is not completely straightforward.

It is a general feature of 5D SUSY GUTs to have a minimal
Higgs sector since the gauge symmetry is, at least in part, broken 
by compactification. Moreover a $U(1)_{R}$ symmetry is naturally present
in 5D constructions and plays the important role of preventing
too fast a proton decay. The allowed superpotential
is rather restricted and the approach of extending it to include
non-minimal terms related to higher Higgs representations 
is not as efficient as in the 4D case.
Instead of dealing with extra Higgs fields, an alternative approach 
is to include extra matter multiplets, such as {\bf 10}-plets of SO(10).
In this way not all of the observed fermions in a given generation 
come from a single {\bf 16}-plet, which gives rise to a much more flexible 
framework. For instance, SU(5)-like lopsided mass matrices for 
down quarks and charged leptons can be constructed and it is possible 
to better exploit flavour symmetries in SO(10) to correctly describe
fermion masses and mixing angles \cite{so10FN}. 

This second approach can be incorporated in a SO(10) 5D construction
and the present note provides a concrete realization
of this idea. The hierarchy between the third generation
and the other two is described by two small parameters, $\epsilon_1$ 
and $\epsilon_2$, arising because the third generation lives on the 
PS brane, while the other two live in the bulk.
The relative suppression between second and third generation,
$\epsilon_2$, is given by the geometrical normalization factor 
$\epsilon=1/\sqrt{\Lambda \pi R}$ of a flat zero mode relative 
to a brane field. The relative suppression between first and third generation,
$\epsilon_1=\delta$, is slightly smaller than $\epsilon$,
and arises from a zero mode with a non trivial profile in $y$.
This setup is tuned to reproduce the mass hierarchy in the up quark sector,
which fixes approximately $\epsilon\approx\lambda^2$ and 
$\delta\approx\lambda^3$.
The down quark and the charged lepton masses are described by
introducing a relative suppression $\epsilon_u$
between $(l,d^c)$ and $(q,u^c,e^c,\nu^c)$,
within the third generation alone. Such a suppression is 
made possible by the fact that these two sectors do not belong
to the same SO(10) irreducible representation. They are effectively 
embedded into a pair $16\oplus 10$, whose extra components acquire
a very large mass. Order of magnitudes are correctly reproduced 
by taking $\epsilon_u\approx\lambda^2$.

Such a construction does not leave too much freedom to the neutrino sector.
The heavy Majorana neutrino mass matrix and the Dirac neutrino mass matrix
have the same order-of-magnitude structures of the up quark mass matrix
and the charged lepton mass matrix, respectively. It is quite 
remarkable that through the see-saw mechanism they gives rise to a successful
neutrino spectrum of normal hierarchy type
with a large atmospheric mixing angle arising from the lopsided
structure of the Dirac sector. 
The whole Yukawa sector is controlled by the VEVs of few
multiplets: $H_{10}$, that breaks the electroweak symmetry,
$\Sigma\oplus\bar{\Sigma}$, that breaks the PS symmetry 
down to the SM one and an SO(10) singlet $\theta$, that 
controls the absolute scale of neutrino masses.
Early works based on 5D SO(10) can be found  in \cite{Shafi5Dso10, Kitano5Dso10}. The authors of \cite{Shafi5Dso10} combine a traditional U(1) flavour symmetry with the SO(10) GUT formulated in 5D.  In this sense, the role of extra dimensions is marginal in order to reproduce their fermion mass pattern. Alternatively, in \cite{Kitano5Dso10}, the fermion mass hierarchy is generated by the breaking of the $U(1)_{X}$ subgroup of SO(10) in the bulk. The effect of this breaking is equivalent to introduce different bulk masses for matter hypermultiplets changing their bulk wave-function profile. However, in order to break  $SO(10) \rightarrow SU(5) \times U(1)_{X}$, they have to introduce a {\bf 45} Higgs representation and, as in the first approach, an additional Dimopoulos-Wilczek mechanism is necessary to provide D-T splitting. 

While proceeding in the construction of the Yukawa sector, our hope
was that the peculiar setup we were defining could manifest in a direct
and observable way at the level of proton decay. For this reason we
have carefully analyzed proton decay in our model. 
There are no renormalizable interactions that violate baryon or lepton number
and dimension five operators due to higgsino exchanged are forbidden.
Baryon violating processes proceed through gauge vector boson exchange and are
described by dimension six operators in the low energy theory.
However, due to the specific localization of matter fields
that emerges from the discussion of Yukawa couplings, there are no
minimal couplings of fermions of first, second and third generations
to the heavy gauge bosons X and Y that mediate proton decay.
Non-minimal interactions are possible and we have listed the 
dominant ones. The main contribution to proton decay is
described by a dimension 6 operator involving fermions of the first 
generation which is suppressed by the cut-off scale $\Lambda$,
which replaces the heavy gauge boson masses $M_X$. As a result, unfortunately,
the possibilities of detecting proton decay, even with future super-massive
detectors, appear to be quite remote.


\vspace*{1.0cm}
{\bf Acknowledgments}
We thank Arthur Hebecker for useful comments. This project is partially
supported by the European Program MRTN-CT-2004-503369.

\end{document}